\documentclass[journal=nalefd,manuscript=article,layout=traditional]{achemso}

\usepackage[version=3]{mhchem} % Formula subscripts using \ce{}
\usepackage{natbib}
\usepackage{color}

\author{T.\,Bathon} 
\affiliation{Physikalisches Institut, Experimentelle Physik II, Universit\"{a}t W\"{u}rzburg, Am Hubland, D-97074 W\"{u}rzburg, Germany}
\author{P.\,Sessi} 
\email{paolo.sessi@physik.uni-wuerzburg.de}
\affiliation{Physikalisches Institut, Experimentelle Physik II, Universit\"{a}t W\"{u}rzburg, Am Hubland, D-97074 W\"{u}rzburg, Germany}
\author{K.\,A.\,Kokh} 
\affiliation{V.S. Sobolev Institute of Geology and Mineralogy, Siberian Branch, Russian Academy of Sciences, 630090 Novosibirsk, Russia}
\affiliation{Novosibirsk State University, 630090 Novosibirsk, Russia}
\author{O.\,E.\,Tereshchenko} 
\affiliation{A.V. Rzanov Institute of Semiconductor Physics, Siberian Branch, Russian Academy of Sciences, 630090 Novosibirsk, Russia}
\affiliation{Novosibirsk State University, 630090 Novosibirsk, Russia}
\author{M.\,Bode}
\affiliation{Physikalisches Institut, Experimentelle Physik II, Universit\"{a}t W\"{u}rzburg, Am Hubland, D-97074 W\"{u}rzburg, Germany}
\affiliation{Wilhelm Conrad R{\"o}ntgen-Center for Complex Material Systems (RCCM), 
Universit\"{a}t W\"{u}rzburg, Am Hubland, D-97074 W\"{u}rzburg, Germany}

\title{Systematics of molecular self-assembled networks \\ at topological insulators surfaces}
\abbreviations{IR,NMR,UV}

\begin{document}

%%%%%%%%%%%%%%%%%%%%%%%%%%%%%%%%%%%%%%%%%%%%%%%%%%%%%
%% The "tocentry" environment can be used to create an entry for the graphical table of contents. 
%% It is given here as some journals require that it is printed as part of the abstract page. 
%% It will be automatically moved as appropriate.
%%%%%%%%%%%%%%%%%%%%%%%%%%%%%%%%%%%%%%%%%%%%%%%%%%%%%
%\begin{tocentry}
%\includegraphics[width=9.0cm]{GraphicalTOC_Entry.pdf}
%\end{tocentry}
\vfill

%%%%%%%%%%%%%%%%%%%%%%%%%%%%%%%%%%%%%%%%%%%%%%%%%%%%%
%% The abstract environment will automatically gobble the contents if an abstract is not used by the target journal.
%%%%%%%%%%%%%%%%%%%%%%%%%%%%%%%%%%%%%%%%%%%%%%%%%%%%%
%\date{\today}
\pagebreak
\begin{abstract}
The success of topological insulators (TI) in creating devices with unique functionalities 
is directly connected to the ability of coupling their helical spin states to well defined perturbations. 
However, up to now, TI-based heterostructures always resulted in very disordered interfaces, 
characterized by strong mesoscopic fluctuations of the chemical potential 
which make the spin-momentum locking ill-defined over length scales of few nanometers 
or even completely destroy topological states.
These limitations call for the ability to control topological interfaces with atomic precision. 
Here, we demonstrate that molecular self-assembly processes driven by inherent interactions among the constituents 
offer the opportunity to create well-defined networks at TIs surfaces. 
Even more remarkably, we show that the symmetry of the overlayer 
can be finely controlled by appropriate chemical modifications. 
By analyzing the influence of the molecules on the TI electronic properties, 
we rationalize our results in terms of the charge redistribution taking place at the interface. 
Overall, our approach offers a precise and fast way to produce tailor-made nanoscale surface landscapes. 
In particular, our findings make organic materials ideal TIs counterparts, since they offer the possibility 
to chemically tune both electronic and magnetic properties within the same family of molecules, 
thereby bringing us a significant step closer towards an application of this fascinating class of materials.
\end{abstract}

Engineering well-ordered nanostructures by using single atoms or molecules 
as building blocks offers a convenient opportunity to control matter at the nanoscale. 
Ultimately, this bottom-up approach represents an alternative 
or may even replace traditional top-down routes like lithography, 
which dominated the electronic industry in the past decades but have recently been challenged 
by the continuos device miniaturization and the related emerging technical difficulties.
The extensive study of the mechanisms driving the creation of self-assembled nanostructures, 
mainly performed in the last decade, resulted in a quite thorough understanding 
of the relevant interactions at the nanoscale and in the development of guidelines 
to create tailor-made nanostructures with extremely high precision~\cite{BCK2005,B2007}.

Although well-defined self-assembled superstructures can be obtained by using both, 
atoms and molecules, the latter are of particular interest~\cite{B2007}. 
This is mainly caused by the fact that molecules are highly controllable by chemical modifications. 
Thereby it becomes possible to tune the molecule's functionality and---at the same time---allowing for 
a large variety of self-assembled nanostructures~\cite{YYK2001,SRA2007,PAS2007,SDK2007,OSM2005,PBP2010}. 
Since the self-assembly process depends on the detailed balance of intermolecular and molecule--substrate interactions, 
there are severe limitations regarding the choice of both, molecules and substrates. 
To date, the most successful ingredients for the emergence of supramolecular order are represented by planar molecules 
coupled to noble metal surfaces~\cite{B2007,YYK2001,SRA2007,PAS2007,SDK2007,OSM2005,PBP2010}, graphite \cite{ASN2007}
and, more recently, to graphene~\cite{MZJ2009,HSD2012,JHB2013} and boron nitride \cite{JHB2013,SDH2013} 
 
Here, we provide evidence that well-ordered molecular superstructures can be obtained 
on the new class of materials named topological insulators (TI)~\cite{HQW2008,CAC2009}. 
Even more remarkably, we demonstrate that both the periodicity and the symmetry of the resulting overlayer can be finely controlled, 
making self-assembly processes a new and versatile way to engineer periodic landscapes on topological states 
without introducing strong mesoscopic fluctuations of the chemical potential 
which make the spin-momentum locking ill defined over length scales of few nanometers 
or even completely destroy topological states\cite{BRS2011,SRB2014}. 
In particular, by using molecules hosting magnetic moments as building blocks, 
the self-assembly process results in the creation of a regular network of localized spins. 
This makes hybrid organic--TI interfaces a promising and reliable platform for the investigation of exotic states of matter 
and the unconventional magneto-electric effects, which have been predicted to exist when topological states 
interact with local spins and which may find direct application 
in devices with new functionalities~\cite{GF2010,M2010,NN2011}.
More generally, our approach identifies planar molecules 
as the ideal TI counterpart as they overcome all the problems 
which have so far limited the study, understanding, and utilization of the response 
of topological states to external perturbations.

As a prototype system, we focus on transition metal (TM)-phthalocyanine (Pc) molecules 
coupled to the chalcogenide topological insulator Bi$_2$Te$_3$. 
Pcs are among the most widely studied planar molecules which are already successfully employed 
in several applications such as sensors and magnets~\cite{H2009}.  
Furthermore, they have been recently reported to constitute a rich playground 
hosting unconventional properties associated with their spin-degree of freedom \cite{FSP2011,HBP2013}.
As shown in Fig.\,1(a), their structure consists of a central metal atom surrounded by an organic macrocycle.
The molecular symmetry is 4-fold and the central atom has a square-planar coordination. 
In TM-Pcs, this bonding scheme leaves the central atom in a $+2$ state and, within the $D_{4h}$ point symmetry, 
the $3d$ levels split into $d_{xz,yz}$, $d_{z^2}$, $d_{xy}$, and $d_{x^2-y^2}$ as illustrated in Fig.\,1(b)~\cite{WSC2009}.
Depending on their occupation, energy, and symmetry, these states can mix with the orbitals of the pyrrole-like rings.
Since the hybridization between $3d$ levels and molecular orbitals may result in several quasi-degenerate configurations, 
it is impossible to draw a definitive picture of the energy occupation level based on a single-electron approach. 
This is particularly true for molecules coupled to substrates, 
since charge transfer processes and modified ligand fields may further complicate the picture.  
However, by increasing the number of $3d$ electrons the following trends emerge: 
(i) the presence, close to the Fermi level, of empty states with $d_{z^2}$ symmetry is gradually reduced;  
(ii) the TM-substrate distance increases~\cite{WSC2009,ZDG2011,MKR2011,MRK2012}.
These observations imply that, contrary to ``late'' transition metal (Ni and Cu) Pc molecules, 
where the central atom's $3d$ orbital does not play any significant role in the adsorption process, 
the situation is quite different for MnPc, FePc, and CoPc.
As described in the following, this has far reaching implications. 

\begin{figure}[t] 
\includegraphics[width=0.9\columnwidth]{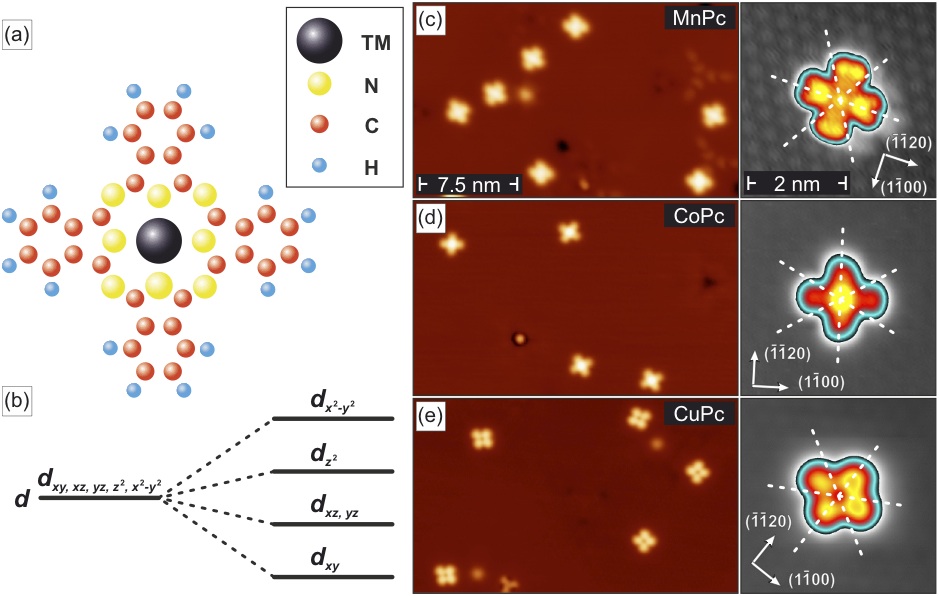}
\caption{(a) Structure of a Pc molecule, which can host several different elements as central atom, 
	offering the opportunity to tune its properties within the same structure. 
	(b)~Crystal field-split energy level for $3d$ orbitals in $D_{4h}$ symmetry. 
	(c-e)~Constant-current images obtained for very dilute concentrations 
	of MnPc, CoPc, and CuPc molecules grown on Bi$_2$Te$_3$, respectively (left panel). 
	Atomically resolved images reveal that, despite the very different electronic configuration of the central $3d$ atom, 
	single MnPc, CoPc, and CuPc molecules all sit on top of a Te atom (right panel). 
	This observation indicates the leading role played by the ligand in determining the adsorption geometry. 
	Note that, while the central atom appears as a protrusion in MnPc and CoPc, 
	it corresponds to a depression in CuPc, thereby indicating the absence of $3d$ orbitals 
	projecting outside the surface plane which can effectively couple to tip states. 
	Scanning parameters: $I = 25$\,pA, $V = 300$\,mV, 500\,mV, and $-200$\,mV for MnPc, CoPc and CuPc, respectively. }  \label{Fig1}
\end{figure}
Figure\,1(c)-(e) display scanning tunneling microscopy (STM) images 
obtained for a very dilute concentration of three different transition metal-phthalocyanine molecules, 
i.e.\ MnPc, CoPc, and CuPc on Bi$_2$Te$_3$, thereby spanning the entire $3d$ occupancy scenario. 
Three relative orientations with respect to the substrate are visible for all three molecules, 
resulting from the combined symmetry of the molecules (4-fold) with that of the substrate (6-fold)~\cite{SBK2014}.  
Inspection of atomically resolved images evidences that, despite their very different $3d$ filling, 
all molecules have the same adsorption geometry, with the central atom sitting on top 
of a Te atom of the underlying Bi$_2$Te$_3$ surface. 
This observation evidences the leading role played by the ligands in determining the adsorption configuration. 
Nevertheless, a direct participation of the central atom 
in the molecule--substrate bond can further anchor the molecules to the surface. 
This is clearly the case for MnPc, whose 2-fold symmetry reduction is indicative 
of a strong molecule--substrate interaction~\cite{CKB2008,CCW2010}. 
In contrast, CoPc and CuPc preserve the 4-fold symmetry typical of the gas phase (see supplementary information). 

The molecule--substrate interaction observed for MnPc is attributed to $d_{z^2}$ orbitals bonding to the TI surface. 
However, as the number of $d$ electrons increases, they shift towards negative energies. 
This is in particular the case for CuPc, where they become double occupied 
without any significant hybridization with the substrate~\cite{ZDG2011,MRK2012}. 
Indeed, while the central atom is imaged as a protrusion on MnPc and CoPc, 
it appears as a depression in CuPc, indicating the absence---in the probed energy range---of $d_{z^2}$ orbitals, 
i.e.\ those electronic states which can effectively couple to the STM tip because of their larger extension into the vacuum 
(for a discussion on the impact of the central atom in STM images see Ref.~\citenum{AEK2015} and references therein). 
The different occupation of the strongly directional $3d$ levels thus offers a convenient way 
to chemically control the delicate balance of forces present at the interface as the molecule coverage increases 
and molecule--molecule interactions start to play an important role. 
Indeed, earlier studies on metallic substrates already pointed out 
the importance of the $d$-filling and the role of the molecule--substrate interaction. 
While an abrupt transition from a dispersive distribution to a square lattice superstructure 
was generally observed on metal substrates \cite{JXL2011,LHW1996,CGD2007}, 
we demonstrate below that on topological insulators an appropriate choice of the central atom 
not only allows to ``activate'' the self-assembly process, but also to select its symmetry.   

\begin{figure}[t] 
\includegraphics[width=0.65\columnwidth]{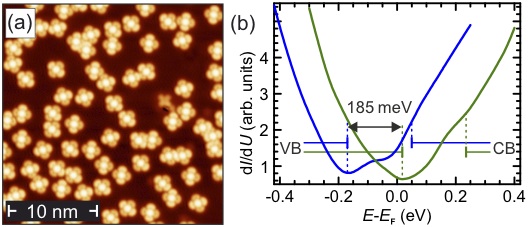}
\caption{(a) Constant-current image obtained for a MnPc concentration equal to 0.13\,molecules/nm$^2$. 
	Despite the small intermolecular distance, cluster formation cannot be observed on the surface, 
	indicating a significant molecule--molecule repulsion. 
	(b)~STS spectra obtained onto the Bi$_2$Te$_3$ before (green line) and after (blue line) molecule deposition. 
	A rigid energy shift towards negative energies appears after the growth. 
	This indicates that the creation of the molecule--TI interface is accompanied by charge transfer processes 
	which $n$-dope the sample leaving MnPc molecules positively charged. 
	Scanning parameters:  $I = 15$\,pA, $V = 500$\,mV. }
\label{Fig2}
\end{figure} 
Figure\,2(a) shows MnPc molecules on Bi$_2$Te$_3$ for a coverage equal to 0.13\,molecules/nm$^2$. 
Note that, despite the reduced molecule--molecule distance, 
not any signature of molecular clustering is present on the surface.  
On the contrary, molecules seem to maximize their distance.
This behavior can be understood by comparing the electronic properties of the Bi$_2$Te$_3$ surface 
before and after molecule deposition as inferred by scanning tunneling spectroscopy (STS).   
Results are reported in Fig.\,2(b).  
The positions of the valence band maximum and conduction band minimum 
have been assigned according to the procedure described in Ref.~\citenum{SOB2013}. 
The spectra evidence a negative rigid shift of the Bi$_2$Te$_3$ band structure 
subsequent to deposition which is indicative of an $n$-doping effect caused by MnPc. 
This charge redistribution leaves the molecules positively charged. 
Consequently, they are stably anchored to the substrate 
by the creation of an interfacial dipole which suppresses their mobility on the surface. 
As a result of the intermolecular repulsive interaction and the relatively strong anchoring to the substrate, 
the molecules do not show any tendency of ordering, irrespective of the coverage.
Furthermore, the second MnPc layer starts to form well before completion of the first one (see Fig.\,3 in Supplementary Information). 
These observations show that repulsive-driven self-assembly processes as observed in Ref.~\citenum{BSA2013}
are absent for the particular combination of molecule and substrate considered here.  

\begin{figure}[t] 
\includegraphics[width=0.99\columnwidth]{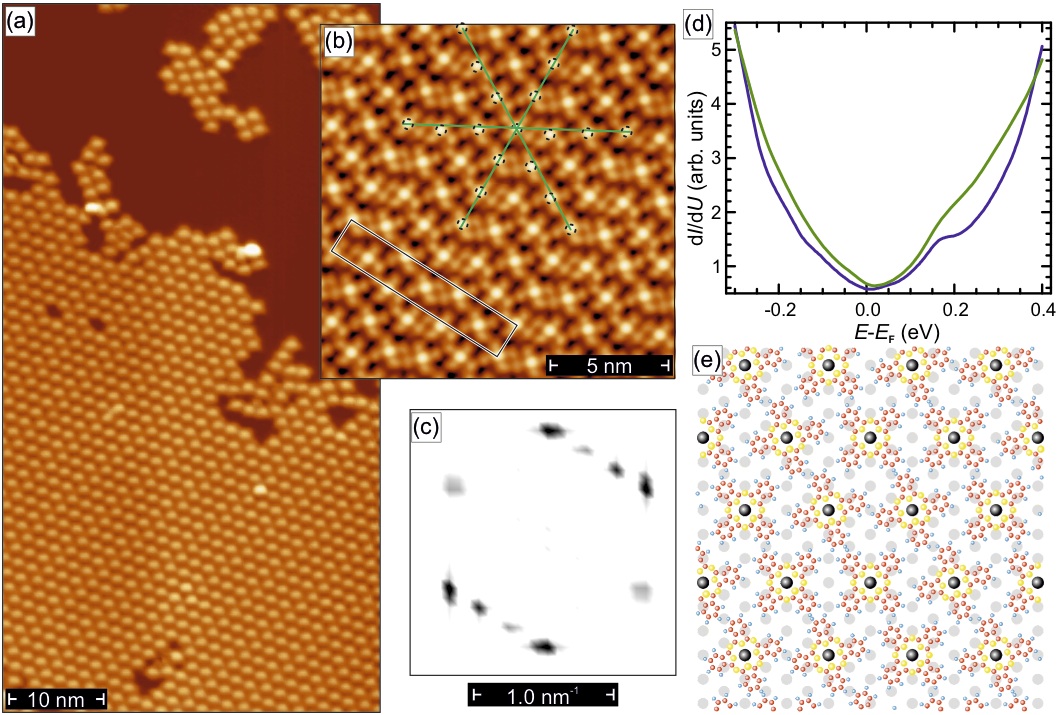}
\caption{(a,b) Self-assembled CoPc molecular film. 
	The unit cell is indicated by a black box.
	(c) Fourier-transformed constant-current image displaying a slightly distorted 
	hexagonal symmetry of the 2D molecular overlayer. 
	(d) In contrast to MnPc [cf.\ Fig.\,2(b)], STS reveals 
	that no significant charge transfer takes place at the interface, 
	indicating that CoPc molecules are weakly bound to the substrate thus retaining substantial surface mobility.
	(e) Model illustrating the structure of the self-assembled molecular film 
	with respect to the underlying Bi$_2$Te$_3$ substrate. 
	Scanning parameters: $I = 15$\,pA, $V = 500$\,mV. }
\label{Fig3}
\end{figure}
A very different scenario appears for CoPc as illustrated in Fig.\,3(a) and (b). 
In this case, large regions displaying a very well-ordered molecular superstructure are present on the surface, 
indicating a relatively weak molecule--substrate interaction as compared to attractive molecule--molecule interactions. 
The weak molecule--substrate bond is also experimentally signaled by significant tip--sample forces, 
which, despite the very low set-point currents used when scanning the surface, 
results in molecules that are occasionally moved by the tip (see Supplementary Information).
As for MnPc, the inspection of spectra obtained before and after deposition allows to visualize 
the charge redistribution processes associated with the creation of the molecule--TI interface. 
Results are reported in Fig.\,3(d).  
The absence of any significant energy shift implies that, contrary to MnPc, CoPc molecules are left in their neutral state. 
This is the more remarkable as the coverage amounts to 0.32\,molecules/nm$^2$, 
i.e.\ well above the one investigated in Fig.\,2 for MnPc.
Contrary to similar close-packed layers obtained at metal surfaces\cite{ASK2010}, 
where close-packed layers evolve but charge-transfer nevertheless occur, 
in the present case molecules have been deposited onto a substrate, 
i.e.\ Bi$_2$Te$_3$, with a much lower density of states. 
As a result, any potential charge transfer taking place at the interface would give rise to a significant band bending, 
as it has been widely reported for topological insulator surfaces 
coupled to single adatoms (see Ref.~\citenum{SBK2014} and~\citenum{VPG2012}) 
or even simply exposed to residual gases (see Refs.~\citenum{BHM2011} and~\citenum{BLK2011}). 
It is worth noticing that the observed difference in charge transfer between MnPc and CoPc 
is further corroborated by simple electronegativity concepts. 
Indeed, as compared to Mn (1.55), the electronegativity of Co (1.88) is closer to those of Bi (2.02) and Te (2.10), 
i.e.\ the atomic species of the underlying substrate.

It thus appears that the combination of CoPc with the Bi$_2$Te$_3$ TI substrate is one of the rare cases 
where the creation of an interface leaves either of the constituents essentially unaltered. 
Therefore, we expect that CoPc behaves very differently 
as compared to other adsorbates which may heavily dope TI surfaces 
and even create new electronic states that are absent on the pristine samples~\cite{BHM2011,BLK2011,VPG2012}. 
Similarly, many TMPc molecules may---once coupled to a substrate---substantially change their electronic properties 
through the interplay of substrate screening and hybridization effects~\cite{MKR2011,MRK2012}. 
In contrast, the electronic properties detected for CoPc on Bi$_2$Te$_3$ 
raises the hope that the gas phase functionality may be preserved. 

The geometrical structure which appears from inspection 
of Fig.\,3(b) allows to identify a quasi-hexagonal crystal lattice, 
which indicates that the symmetry of the substrate has been transferred to the overlayer. 
Minor periodic distortions are visible along two of the three high symmetry axis, 
also signaled by the appearance of 4 additional spots in the Fourier transformation [see Fig.\,3(c)]. 
Although the presence of a well-ordered superstructure indicates that molecule--molecule interactions dominate 
the delicate balance of forces that drive the self-assembly process, molecule--substrate interactions 
still appear to play an important role since it determines the axis along which molecules self-assemble.
A detailed analysis of the molecular film allows to obtain the structural model illustrated in Fig.\,3(e).  
Molecules have a nearest-neighbor distance corresponding to 1.55~nm, 
with each of them sitting on top (or in very close proximity, see below) of the underlying Te atom, 
as can also be inferred from Fig.\,3(b), where all molecules display the very same appearance. 
Furthermore, the different rotations of the molecules allow to explain the distortions introduced in the lattice 
as the result of small steric repulsion-induced deviations with respect to the minimum energy adsorption site. 
Overall, these observations point towards an overlayer unit cell that is commensurate with the substrate.

Since MnPc and CoPc have the same structure and ligand, their different behavior in the high coverage regime 
must be a direct consequence of the different electronic configuration of their $3d$ central atoms.
Although a precise analysis of the molecule--substrate interactions would require detailed calculations 
going beyond the scope of the present work, an heuristic picture that is based on 
the $3d$ levels occupancy can effectively explain our findings. 
In particular, the differences are ascribed to the presence of $d_{z^2}$ orbitals close to Fermi level. 
These electronic states effectively facilitate the bonding to the substrate, 
which is gradually reduced as the number of electrons residing on the central atom increases, 
i.e.\ by changing from a Mn to a Co ion~\cite{WSC2009}. 
Therefore, changing to a higher atomic number drives the balance 
of molecule--substrate and molecule--molecule interactions towards the latter.   

\begin{figure}[t] 
\includegraphics[width=0.99\columnwidth]{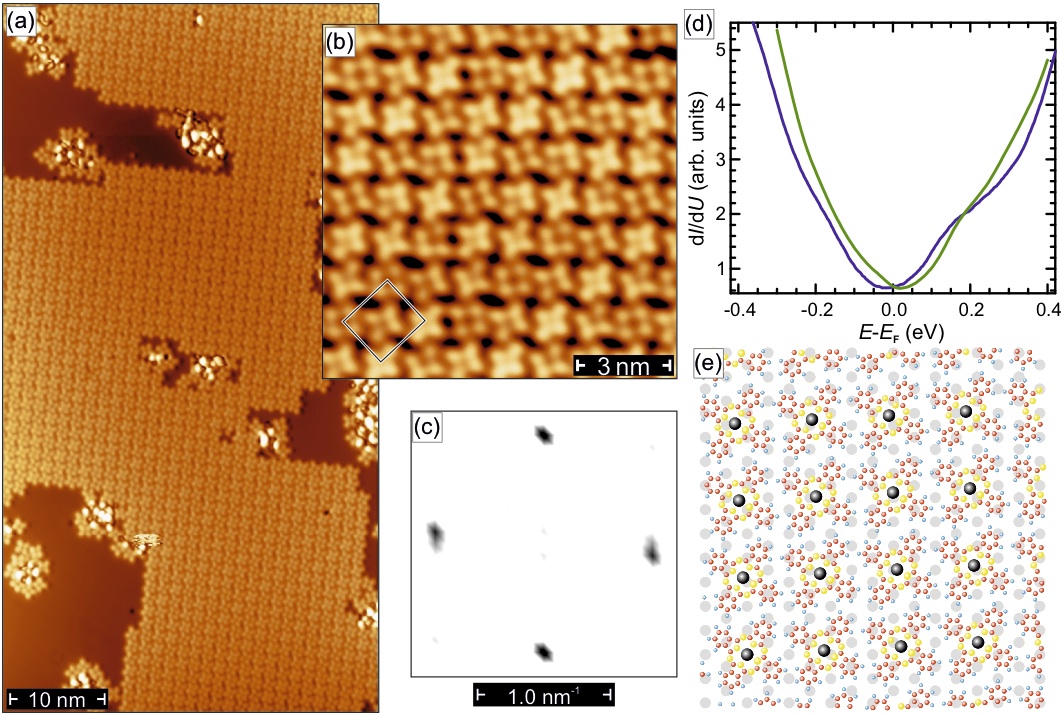}
\caption{(a) Self-assembled CuPc molecular film. 
	(b) Contrary to CoPc [cf.\ Fig.\,3(b)], CuPc self-assembles 
	with a square symmetry (unit cell indicated by box), 
	as evidenced by the Fourier transformed constant-current image displayed in (c).  
	(d) Similar to CoPc, STS reveals that no significant charge transfer takes place at the interface, 
	indicating that CuPc molecules are weakly bound to the substrate.
	(e) Model illustrating the structure of the self-assembled molecular film 
	with respect to the underlying Bi$_2$Te$_3$ substrate. 
	Scanning parameters:   $I = 15$\,pA, $V = 500$\,mV.}
\label{Fig4}
\end{figure}
To test our model, measurements have been performed on CuPc which, 
because of its double-occupied $d_{z^2}$ orbital 
that lies well below the Fermi level~\cite{ZDG2011,MRK2012}, 
is expected to result in an even weaker molecule--substrate interaction with respect to CoPc.  
By looking at Fig.\,4(a) it is evident that a well-ordered self-assembled structure 
is obtained also in this case (coverage 0.46\,molecules/nm$^2$).  
As for CoPc, the creation of a molecular film is not associated 
with any significant charge transfer between molecules and substrate [see Fig.\,4(d)] 
[note that, as in the case of CoPc (see above), also for CuPc a simple electronegativity picture 
supports our findings, being Cu electronegativity equal to 1.90]. 
However, a zoomed-in image demonstrates that, contrary to CoPc, 
CuPc self-assembles in a closed-packed structure with a square lattice [see Fig.\,4b]. 
This is the typical lattice adopted by Pc molecules deposited on weakly interacting substrates, 
since it allows to maximize their mutual interaction.
Indeed, the nearest-neighbor distance of 1.34~nm is essentially determined by the size of the molecule. 

An analysis of the molecular film allows to obtain the structural model reported in Fig.\,4(e). 
Since the symmetries of the molecular film and the substrate are incompatible, 
the molecules adopt different adsorption configurations 
with respect to the underlying Bi$_2$Te$_3$ surface, 
as also signaled by their different appearance in Fig.\,4(b).  
The long wavelength modulation visible in the image suggests an incommensurability 
between the molecular layer and the substrate, as already discussed in Ref.\,\citenum{ASK2010}. 
Unambiguous determination would require simultaneous atomic resolution of the underlying substrate 
which was impossible to achieve because of incompatible tunneling parameters.
Overall, these observations are in agreement with our model 
and confirm the leading role played by the central $3d$ orbitals 
in balancing the different interaction taking place at the molecule--TI interface.

In summary, our findings show that appropriate chemical modifications of metal-organic compounds 
lead to a rich variety of hybrid molecule-TI interfaces.
As a result these interfaces represent ideal heterostructures 
for tailoring potential landscapes at TI surfaces at the atomic scale. 
Furthermore, molecules with built-in magnetic moments, such as the molecules used in this study,  
may result in superstructures that constitute well defined spin networks, 
potentially leading to device concepts with unique functionalities.

{\bf Methods}
Experiments have been performed in a commercial STM operated at $T = 4.8$~K. 
The Bi$_2$Te$_3$ single-crystal samples were synthesized 
by mixing stoichiometric amounts of bismuth and tellurium. 
The crystal structure consists of alternating planes of Bi and Te 
up to the formation of a quintuple layer with the sequence Te-Bi-Te-Bi-Te. 
Quintuple layers are weakly coupled by van der Waals forces thus offering a natural cleaving plane. 
Bi$_2$Te$_3$ single crystals have been cleaved in UHV 
at a base pressure of $3 \cdot 10^{-11}$ mbar and immediately inserted into the STM. 
MnPc, CoPc and CuPc molecules (Sigma-Aldrich) were deposited directly onto the cold Bi$_2$Te$_3$ surface 
by using a home made Knudsen cell and annealed at room temperature. 
STM measurements were performed using electrochemically etched tungsten tips. 
Topographic images were acquired in the constant-current mode. 
Spectroscopic data were obtained by lock-in technique ($f = 793$~Hz, $V_{rms} = 10$~mV). 
Since the apparent molecular coverage was found to strongly depend on the particular bias voltage applied, 
an effect that is related to the spatial distribution of certain molecular orbitals, 
coverages are given in number of molecules per nm$^2$. \\
 	
{\noindent \bf Associated Content}\\
{\sc Supporting information}\\
Description of the adsorption of TM-Pc on Bi$_2$Te$_3$ which lead to a C4 to C2 symmetry reduction; 
evidence for weak adsorption energy from strong tip-molecule interaction; 
a coverage-dependent study of MnPc molecules; 
and the cluster formation at very dilute concentration regime for CoPc and CoPc.  
This material is available free of charge via the Internet at http://pubs.acs.org.

\begin{acknowledgement}
This work has been funded within SPP 1666 ``Topologische Isolatoren'' (project BO 1468/21-1).
K.A.K. and O.E.T. acknowledge financial support by the RFBR (Grant Nos.\ 13-02-92105 and 14-08-31110).
\end{acknowledgement}

\end{document}